\documentclass[prb,aps,superscriptaddress,reprint,twocolumn]{revtex4-1}

\usepackage{graphicx}
\usepackage{amsmath}
\usepackage{txfonts}
\begin{document}

\title{Orbital Optimization in the Active Space Decomposition Model}
\author{Inkoo Kim}
\affiliation{\mbox{Department of Chemistry, Northwestern University, 2145 Sheridan Rd., Evanston, IL 60208, USA.}}
\author{Shane M. Parker}
\affiliation{\mbox{Department of Chemistry, Northwestern University, 2145 Sheridan Rd., Evanston, IL 60208, USA.}}
\altaffiliation[Present Address:]{Department of Chemistry, University of California, Irvine, 1102 Natural Sciences II, Irvine, CA 92697, USA.}
\author{Toru Shiozaki}
\affiliation{\mbox{Department of Chemistry, Northwestern University, 2145 Sheridan Rd., Evanston, IL 60208, USA.}}
\email{shiozaki@northwestern.edu}
\date{\today}

\begin{abstract}
\vspace*{-7pt}
We report the derivation and implementation of orbital optimization algorithms for the active space decomposition (ASD) model,
which are extensions of complete active space self-consistent field (CASSCF) and its occupation-restricted variants in the conventional multiconfiguration electronic-structure theory.
Orbital rotations between active subspaces are included in the optimization, which allows us to unambiguously partition the active space into subspaces,
enabling application of ASD to electron and exciton dynamics in covalently linked chromophores. 
One- and two-particle reduced density matrices, which are required for evaluation of orbital gradient and approximate Hessian elements, are computed from the intermediate tensors in the ASD energy evaluation.
Numerical results on 4-(2-naphthylmethyl)-benzaldehyde and [$3_6$]cyclophane
and model Hamiltonian analyses of triplet energy transfer processes in the Closs systems are presented.
Furthermore model Hamiltonians for hole and electron transfer processes in {\it anti}-[2.2](1,4)pentacenophane
are studied using an occupation-restricted variant.
\end{abstract}

\maketitle

\section{Introduction}

Electron and exciton dynamics between chromophores are ubiquitously found---from photosynthetic complexes in biological organisms\cite{Amerongenbook} to organic semiconductors for solar energy conversion.\cite{Desiraju2013CR}
They are often modeled by quantum dynamics methods based on the quantum master equation,\cite{Nitzanbook}
which requires so-called model Hamiltonians in the diabatic representation as an input. 
The model Hamiltonians are the reduced dimensional representation of the electronic degree of freedom in such systems,
consisting of the energies of the diabatic states and the interaction strengths between these states.
Although accurate model Hamiltonians are of essential importance in achieving predictive simulations of electron and exciton dynamics,
there have been only a few studies to develop accurate electronic structure methods to compute them from first principles.\cite{Difley2011JCTC,Kaduk2012CR,Subotnik2010JPCA, Alguire2014JPCA, Landry2014JCTC, Fink2008CP, Havenith2012MP, Wu2011ChemRev}

We have recently introduced the active space decomposition (ASD) method to provide accurate model Hamiltonians.
The ASD model takes advantage of molecular geometries to compress active-space wave functions of molecular dimers,\cite{Parker2013JCP,Parker2014JCTC,Parker2014JPCC,Parker2014JCP}
in which wave functions are parameterized as\cite{Parker2013JCP}
	\begin{align}
	|\Psi \rangle = \sum_{IJ} U_{IJ} |\Phi_I^A\rangle |\Phi_J^B\rangle, \label{eq:wf}
	\end{align}
where $A$ and $B$ label monomers and $I$ and $J$ label monomer states. 
The monomer wave functions ($|\Phi_I^A\rangle$ and $|\Phi_J^B\rangle$)
are determined by diagonalizing the respective monomer active space Hamiltonians (i.e., $\hat{H}^A | \Phi^A_I \rangle = E^A_I | \Phi^A_I \rangle$) in an orthogonal active subspace (hereafter referred to as an ASD subspace).
The expansion in Eq.~\eqref{eq:wf} is exact when each $I$ and $J$ entirely span the corresponding monomer space, and converges rapidly with respect to the number of monomer states included in the summation.\cite{Parker2013JCP}
An extension to more than two active subspaces has also been reported by the authors.\cite{Parker2014JCP}
Furthermore, the dimer basis states have well-defined charge, spin, and spin-projection quantum numbers on each monomer,
allowing us to extract model Hamiltonians for electron and exciton dynamics through diagonalization of diabatic subblocks of a dimer Hamiltonian matrix.\cite{Parker2014JCTC}
This methodology has been applied to singlet exciton fission dynamics in tetracene and pentacene crystals.\cite{Parker2014JPCC}
The resulting model Hamiltonians have been used to benchmark approximate methods in more recent studies.\cite{Alguire2015JPCA} 

However, our ASD method was not applicable to covalently linked chromophores in previous works.
This was mainly due to the difficulty in finding appropriate ASD subspaces. 
In the original formulation of ASD, one must define localized ASD subspaces from localized Hartree--Fock orbitals.\cite{Parker2014JPCC}
Although well defined (and even automated) for molecular dimers and aggregates, this procedure gives rise to an ambiguity in the definition of the ASD subspaces in covalently linked chromophores owing to orbital mixing.
This problem has hampered applications of ASD to these systems.

In this study, we remove this ambiguity in the subspace preparation by deriving and implementing orbital optimization algorithms for the ASD model.
The orbital optimization guarantees that (when converged) identical results are obtained regardless of the choice of initial active orbitals.
As shown in the following, the orbital optimization procedure naturally leads to localized ASD subspaces without deteriorating the diabatic structure of the dimer basis function in the original ASD method.
The preservation of locality of the ASD subspaces during orbital optimization is analogous to the fact that, for a fixed number of renormalized states, {\it ab initio} density matrix renormalization group algorithms using localized orbitals provide the lowest energies. \cite{Chan2011ARPC}
A quasi-second-order algorithm is employed,\cite{Chaban1997TCA} which is akin to conventional complete active space self-consistent field (CASSCF) and its occupation restricted variants (RASSCF).\cite{Olsen1988JCP}
In the following our orbital-optimized models are referred to as ASD-CASSCF and ASD-RASSCF depending on the underlying monomer active-space wave functions.

\section{Theory}
\subsection{Orbital optimization model\label{theorysec}}
In ASD the energy of a dimer state is a function of molecular orbital (MO) coefficients ($\mathbf{C}$), configuration interaction (CI) coefficients within each monomer ($c^X_D$ with $D$ labeling Slater determinants for $X=A$, $B$), and ASD coefficients [$U_{IJ}$ in Eq.~\eqref{eq:wf}], i.e., 
\begin{align}
E_\mathrm{ASD} = E(\mathbf{C}, c^A_D, c^B_{D'}, U_{IJ}). 
\end{align}
We first form the monomer basis by solving configuration interaction within each ASD subspace 
\begin{align}
&\hat{H}^A |\Phi_I^A \rangle = E^A_I |\Phi_I^A \rangle,\\ 
&\hat{H}^B |\Phi_J^B \rangle = E^B_J |\Phi_J^B \rangle,
\end{align}
where $|\Phi_I^A\rangle = \sum_D c_D^{A,I} |D \rangle$ and so on.
Using the monomer basis states, the dimer Hamiltonian matrix elements are computed as
\begin{align}
&\langle \Phi_I^A \Phi_J^B | \hat{H} | \Phi_{I'}^A \Phi_{J'}^B \rangle
= (-1)^\phi\sum_{\zeta\eta}\Gamma^{A,II'}_\zeta h_{\zeta,\eta} \Gamma^{B,JJ'}_\eta,\label{eq:H}
\end{align}
which can be evaluated without constructing the product basis functions explicitly.
Here the indices $\zeta$ and $\eta$ are operators acting on monomer $A$ and $B$, respectively, whose product corresponds to either one- or two-electron operator of the rearranged Hamiltonian; $h_{\zeta,\eta}$ is the molecular orbital integral;
and, $(-1)^\phi$ is a phase factor due to the rearrangement.
The monomer intermediates $\Gamma^{A,II'}_\zeta$ and $\Gamma^{B,JJ'}_\eta$ are defined by
\begin{align}
&\Gamma^{A,II'}_\zeta = \langle \Phi_I^A|\hat{E}_\zeta |\Phi_{I'}^A\rangle, \label{eq:gamma_A} \\
&\Gamma^{B,JJ'}_\eta = \langle \Phi_J^B|\hat{E}_\eta |\Phi_{J'}^B\rangle.\label{eq:gamma_B}
\end{align}
Diagonalization of the dimer Hamiltonian matrix gives ASD energies and wave functions.
The details of the algorithm are given elsewhere.\cite{Parker2013JCP} 

\begin{figure}
\includegraphics[width=0.48\textwidth]{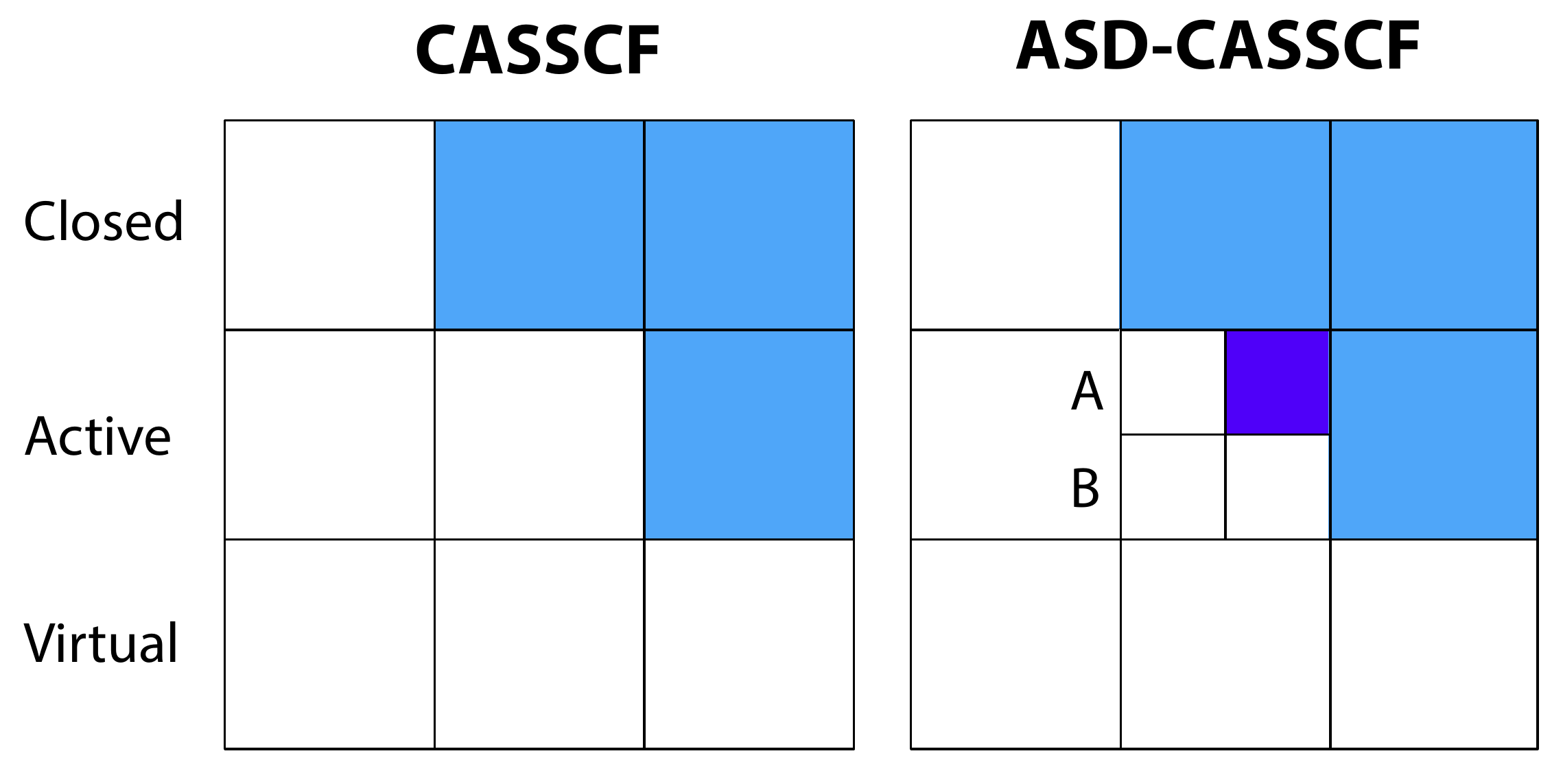}
\caption{\label{fig:rotmat} Non-redundant elements of the rotation matrices $\kappa$ for CASSCF (left) and ASD-CASSCF (right).  }
\end{figure}

We introduce a stationarity condition with respect to variations of MO coefficients to define orbital-optimized ASD models:
\begin{align}
\frac{\partial E_\mathrm{ASD}}{\partial {\kappa}_{rs}} = 0, \label{deriv}
\end{align}
where $r$ and $s$ label any molecular orbitals, and $\boldsymbol{\kappa}$ is the anti-symmetric matrix that parameterizes the MO coefficients,
\begin{align}
\mathbf{C} = \mathbf{C}_\mathrm{init} \exp(\boldsymbol{\kappa}).
\end{align}
The non-redundant elements of the rotation matrix $\boldsymbol{\kappa}$ are depicted in Fig.~\ref{fig:rotmat}.
The two-step optimization procedure (i.e., alternating solutions of ASD and orbital updates until convergence) gives a unique energy that is close to the variationally optimal one.
This quasi-variational behavior is explained as follows.
The total derivatives of the ASD energy with respect to variations of MO coefficients are
\begin{align}
\frac{dE_\mathrm{ASD}}{d {\kappa}_{rs}} &= \frac{\partial E_\mathrm{ASD}}{\partial {\kappa}_{rs}} + \sum_{IJ} \frac{\partial E_\mathrm{ASD}}{\partial U_{IJ}} \frac{d U_{IJ}}{d \kappa_{rs}} \nonumber\\
&+ \sum_{D,I} \frac{\partial E_\mathrm{ASD}}{\partial c^{A,I}_{D}} \frac{d c^{A,I}_{D}}{d \kappa_{rs}}
 + \sum_{D',J} \frac{\partial E_\mathrm{ASD}}{\partial c^{B,J}_{D'}} \frac{d c^{B,J}_{D'}}{d \kappa_{rs}},
\label{totalderiv}
\end{align}
where respective normalization conditions are implicit.
First, it follows from the ASD procedure that the stationary condition with respect to $U_{IJ}$,
\begin{align}
\frac{\partial E_\mathrm{ASD}}{\partial U_{IJ}} = 0, \label{uderiv}
\end{align}
is automatically satisfied.
Second, the following equations are approximately satisfied,
\begin{align}
\frac{\partial E_\mathrm{ASD}}{\partial c^A_D} \approx 0,\quad
\frac{\partial E_\mathrm{ASD}}{\partial c^B_{D'}} \approx 0.
\end{align}
The residuals of these derivatives are related to the differences between the ASD energies computed by lowest-energy monomer states and variationally optimal ones (with the same number of states in the summation),
which are expected to be much smaller than the energies themselves in the cases where ASD is a good approximation. 
As a result, one obtains that the total derivative [Eq.~\eqref{totalderiv}] is approximately zero, making the orbitals almost variationally optimal.
In the cases where convergence is not achieved under this approximation,
we fix $c^A_D$ and $c^B_{D'}$ after several iterations (but not $U_{IJ}$);
when $c^A_D$ and $c^B_{D'}$ are fixed, the last two terms in Eq.~\eqref{totalderiv} that introduce numerical noise become exactly zero because
the derivatives of $c^A_D$ and $c^B_{D'}$ are zero, and hence, the minimum in this constrained frame
can be found by standard minimization algorithms using the partial derivative of energy with respect to orbital rotations [i.e., Eq.~\eqref{deriv}]. 
After the convergence ASD-CI is performed once again using the optimized orbitals. 
The solutions are obtained by iterating this procedure until
consistency between orbitals and $c^A_D$ and $c^B_{D'}$ is achieved.

Lastly, we note that the orbital-optimized ASD energy becomes equivalent to the conventional CASSCF and RASSCF energy when all the charge and spin sectors are included in the ASD expansion.
In Sec.~\ref{results} we show that the energy converges rapidly with respect to the number of states in the ASD expansion.

\subsection{Orbital gradients and approximate Hessians}
In addition to orbital gradients with respect to rotations between closed, active, and virtual subspaces for which gradient elements are known,\cite{Chaban1997TCA}
we consider rotations among ASD subspaces $A$ and $B$.
The energy gradients between orbitals in active space $A$ (unbarred) and those in active space $B$ (barred) are
\begin{align}
& \frac{\partial E_\mathrm{ASD}}{\partial \kappa_{\bar{i}j}} = 2(F_{j\bar{i}}-F_{\bar{i}j}), \label{eq:grad} \\
& F_{\bar{i}j} = \sum_k \Gamma_{jk} F_{\bar{i}k}^I + \sum_{klm} \Gamma_{jk,lm} (\bar{i}k|lm),\\
& F_{\bar{i}j}^I = h_{\bar{i}j} + \sum_{o}[2(\bar{i}j|oo)-(\bar{i}o|oj)],
\end{align}
where $i$, $j$, $k$, $l$, and $m$ label active orbitals (the summation indices $k$, $l$, and $m$ run over orbitals on both $A$ and $B$), and $o$ labels closed orbitals.
$\Gamma_{ij}$ and $\Gamma_{ij,kl}$ are the standard spin-free one- and two-particle reduced density matrices, respectively.

In the quasi-second-order optimization algorithm\cite{Chaban1997TCA} approximate Hessian elements are required as an initial guess.
We use the following formula for approximate diagonal Hessian elements for the inter-subspace active--active rotations,
\begin{align}
\frac{\partial^2 E_\mathrm{ASD}}{\partial \kappa_{\bar{i}j} \partial \kappa_{\bar{i}j}}  &\approx 
2 [ \Gamma_{\bar{i}\bar{i}} ( F^I_{jj} + F^A_{jj} ) + \Gamma_{jj} ( F^I_{\bar{i}\bar{i}} + F^A_{\bar{i}\bar{i}} ) \nonumber \\ 
& -F_{\bar{i}\bar{i}} - F_{jj} - 2 \Gamma_{\bar{i}j} F^I_{\bar{i}j} ],
\end{align}
where $F^A_{ij} = \sum_{kl} \Gamma_{kl}[(ij|kl)-\frac{1}{2}(il|kj)]$.
This formula is derived using similar approximations to those in Ref.~\onlinecite{Chaban1997TCA}, in which 
the two-electron integrals are approximated by the diagonal element of Fock-like one-electron operators and
active orbitals is approximated to be completely filled for some terms.
This procedure eliminates the needs for the computation of two-electron integrals with more than one active indices.

One- and two-particle reduced density matrices in the ASD model, $\Gamma_{ij}$ and $\Gamma_{ij,kl}$, are computed from
$\Gamma_\zeta$ intermediate tensors as using Eqs.~\eqref{eq:wf}, \eqref{eq:gamma_A}, and \eqref{eq:gamma_B} as follows,
\begin{align}
&\Gamma_{\zeta\eta} = \sum_{II'} \Gamma^{A,II'}_\zeta \tilde{\Gamma}^{II'}_{\eta}, \\
&\tilde{\Gamma}^{II'}_{\eta} = \sum_{JJ'}U_{IJ}U_{I'J'}\Gamma^{B,JJ'}_\eta,
\end{align} 
which after index reordering give $\Gamma_{ij}$ and $\Gamma_{ij,kl}$.
Because $\Gamma_\zeta$ intermediates are already computed in the ASD energy evaluation, additional costs for computing density matrices are negligible. 
Note that density matrices have permutation symmetry; for instance,
\begin{align}
\Gamma_{\bar{i}j,kl} = \Gamma_{j\bar{i},lk} = \Gamma_{kl,\bar{i}j} = \Gamma_{lk,j\bar{i}}.
\end{align}
The density matrix elements with all indices belonging to the same monomer are computed separately for computational efficiency as
\begin{align}
& \Gamma_{ij,kl} = \sum_{J} \sum_{\rho\sigma}\langle \tilde{\Phi}^A_J | i^\dagger_\rho k^\dagger_\sigma l_\sigma j_\rho |\tilde{\Phi}^A_J \rangle,\\
& \Gamma_{\bar{i}\bar{j},\bar{k}\bar{l}} = \sum_{I} \sum_{\rho\sigma}\langle \tilde{\Phi}^B_I | \bar{i}^\dagger_\rho \bar{k}^\dagger_\sigma \bar{l}_\sigma \bar{j}_\rho |\tilde{\Phi}^B_I \rangle,
\end{align}
using rotated monomer states $|\tilde{\Phi}^A_J\rangle = \sum_I U_{IJ}|\Phi_I^A\rangle$ and $|\tilde{\Phi}^B_I\rangle = \sum_J U_{IJ}|\Phi_J^B\rangle$.
The use of rotated monomer states reduces the cost of calculating these elements from quadratic to linear scaling with respect to the number of monomer states.

\subsection{State averaging and model Hamiltonians}
As in traditional CASSCF, state averaging is used to optimize orbitals for systems that involve multiple electronic states,
in which $E_\mathrm{ASD}$ in Eq.~\eqref{deriv} is related by an averaged energies of states of interest (labeled by $K$),
\begin{align}
E_\mathrm{ASD}^\mathrm{ave} = \frac{1}{n}\sum_{K=1}^n E^K_\mathrm{ASD}.
\end{align}
This means that all the density matrices $\Gamma_{ij}$ and $\Gamma_{ij,kl}$ that appear in the orbital gradient and approximate Hessian elements
have to be replaced by state-averaged counterparts,
\begin{align}
&\Gamma_{ij}^\mathrm{ave} = \frac{1}{n}\sum_{K=1}^n \Gamma^K_{ij},\\ 
&\Gamma_{ij,kl}^\mathrm{ave} = \frac{1}{n}\sum_{K=1}^n \Gamma^K_{ij,kl}.
\end{align}
The rest of the algorithm remains identical.

As discussed in the introduction, ASD-CASSCF is developed to realize calculation of diabatic model Hamiltonians for electron and exciton dynamics. 
The procedure to extract model Hamiltonians from ASD dimer Hamiltonian matrices has been detailed in Ref.~\onlinecite{Parker2014JCTC}.
In essence, we form a diabatic model state by diagonalizing each diagonal-subblock of the dimer Hamiltonian corresponding to appropriate charge and spin quantum numbers.
The diabatic wave functions are written as 
\begin{align}
| \Psi^K_{\Omega_A\Omega_B} \rangle = \sum_{I\in \Omega_A J\in \Omega_B} U^K_{IJ} | \Phi^A_{I} \rangle | \Phi^B_{J} \rangle,
\end{align}
in which $\Omega_A$ and $\Omega_B$ are collections of monomer states with given charge and spin quantum numbers.
The diagonal elements are the eigenvalues in the subblock diagonalization.
The off-diagonal Hamiltonian matrix element of the diabatic states are defined as
\begin{align}
H_{KK'} = \langle \Psi^K_{\Omega_A \Omega_B} | \hat{H} | \Psi^{K'}_{\Omega'_A \Omega'_B} \rangle. \label{eq:dc}
\end{align}
When computing model Hamiltonians with orbital optimization, it is convenient to use the averaged energy of the diabatic states in the model Hamiltonian.
This procedure gives identical energies to the state-averaged calculations with the adiabatic states obtained by diagonalizing the model Hamiltonian, since
the trace of a Hamiltonian matrix is invariant under the unitary transformation.

\section{Results and discussion\label{results}}

\begin{figure}
\includegraphics[width=0.4\textwidth]{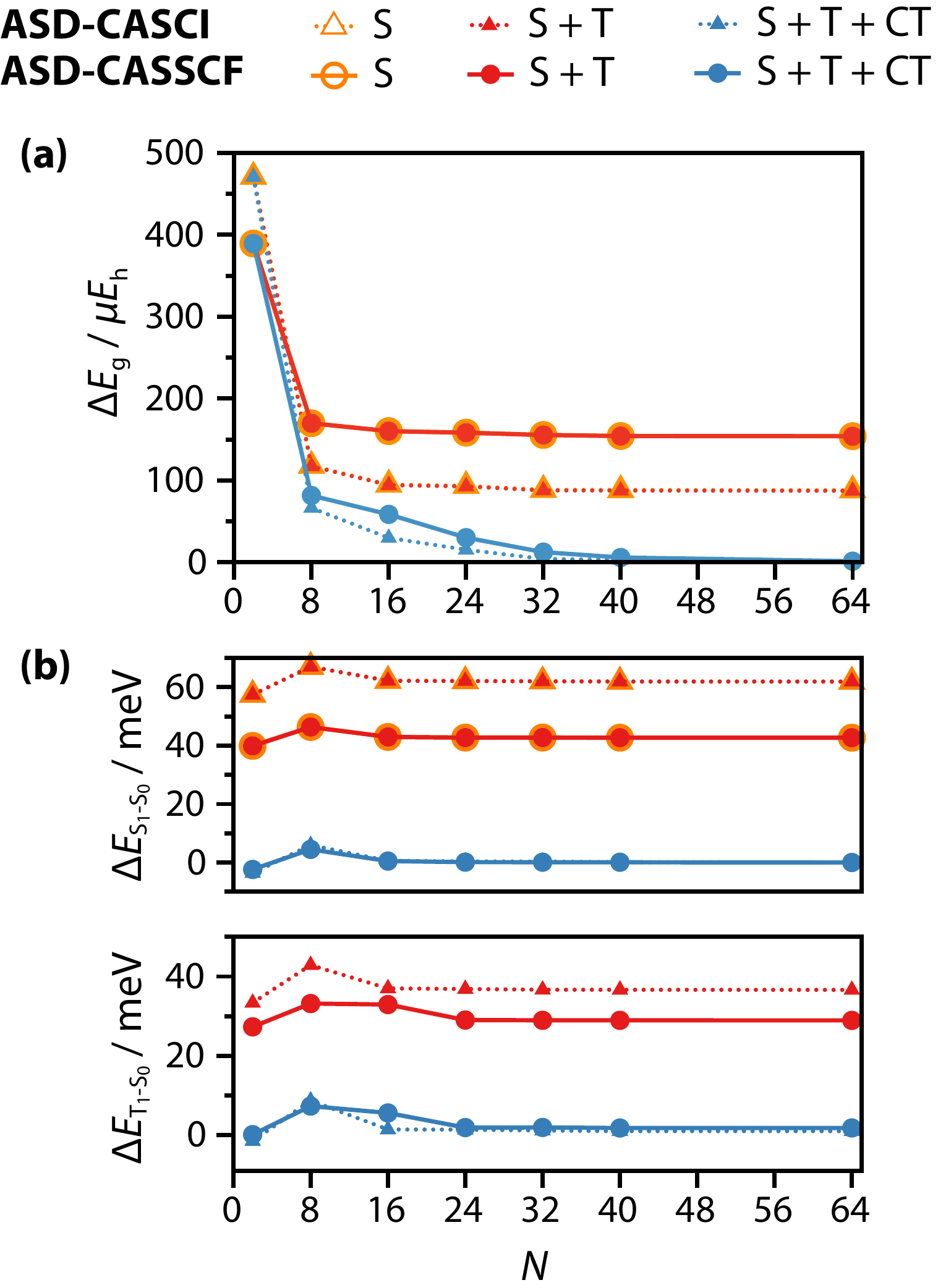}
\caption{\label{fig:bz} 
Errors in (a) the ground-state energy and (b) the energy splitting between the ground and first excited singlet and triplet states of a benzene dimer
as functions of the number of states in each charge and spin sector in the ASD expansion ($N$) computed by ASD-CASCI and ASD-CASSCF.
}
\end{figure}

We applied ASD-CASSCF to the benzene dimer in its sandwich configuration with a separation of 4.0 \AA.
The cc-pVDZ basis set\cite{Dunning1989JCP} was used, and singlet (S), triplet (T), and charge-transfer (CT) monomer states are included.
The CAS(12,12) active space for a dimer consists of 12 $\pi$ electrons distributed in the $\pi$ orbitals,
which is decomposed to two 6-orbital ASD subspaces.
First, the ground state energies and the first few excitation energies computed by ASD-CASCI and ASD-CASSCF as a function of the number of states in each charge and spin sector in the ASD expansion ($N$) are presented in Fig.~\ref{fig:bz}.
The errors are calculated relative to the conventional CASCI ($-461.547119\,E_\mathrm{h}$) and CASSCF ($-461.584702\,E_\mathrm{h}$) energies.
ASD-CASSCF and ASD-CASCI converge to the conventional CASSCF and CASCI with full CAS(12,12) at large $N$.
The Hartree--Fock orbitals were used in the CASCI calculations.
For the ground state, the convergence of ASD-CASSCF is slightly slower than that with ASD-CASCI, although they are both almost exponential.
The energies converged to 0.1 m$E_\text{H}$ with as small as $N=8$. 
We also note that ASD-CASSCF with a more diffuse aug-cc-pVDZ basis set gave almost identical errors for given $N$.
See the supporting information for details.\cite{supp}
The excitation energies show negligible variance with respect to the number of monomer states. 
In all cases, the inclusion of CT states is essential to obtain numerically near-exact energies.
The T-T contributions were negligible for the ground state, since the configuration in which the two monomers have anti-parallel triplets lies too high in energy to interact with the neutral S-S configuration.
Orbital optimization increased and decreased the CT contributions to the ground and excited states, respectively.

\begin{figure}
\includegraphics[width=0.35\textwidth]{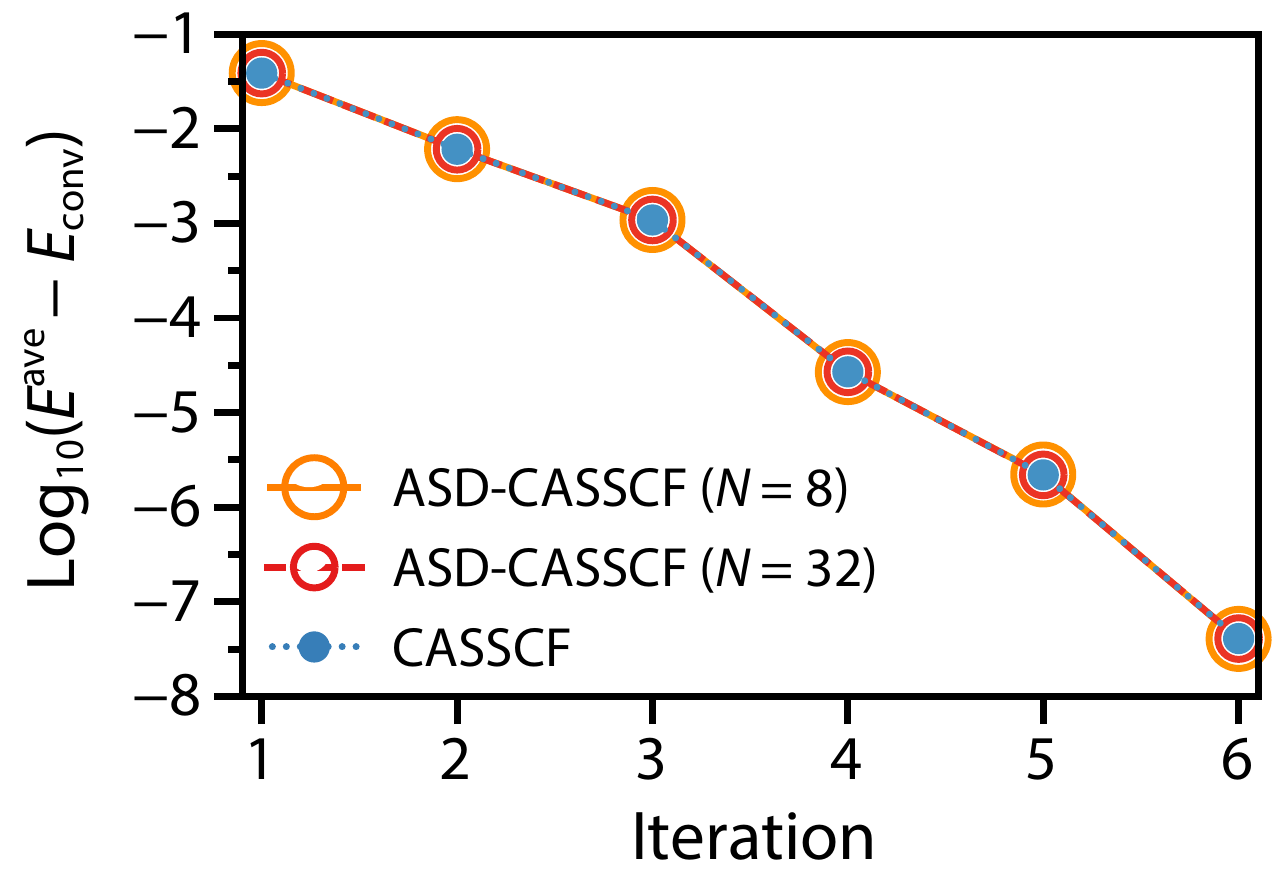}
\caption{\label{conver} 
Convergence of orbital optimization in ASD-CASSCF and the standard CASSCF for a benzene dimer (see text for computational details).
A quasi-second-order algorithm was used.} 
\end{figure}

The convergence of orbital optimization in ASD-CASSCF is compared to that in CASSCF and is shown in Fig.~\ref{conver}.
Three states were state-averaged in the calculations.
We found that the convergence behaviors of ASD-CASSCF and CASSCF were nearly identical,
corroborating that active--active rotations do not deteriorate the numerical stability of orbital optimization.
The total ASD-CASSCF calculation (7 iterations) took about 75 and 210 seconds with $N=8$ and $32$,
while the standard CASSCF took about 3300 seconds
using 16 CPU cores of Xeon E5-2650 2.00~GHz.

\begin{figure}[t]
\includegraphics[width=0.37\textwidth]{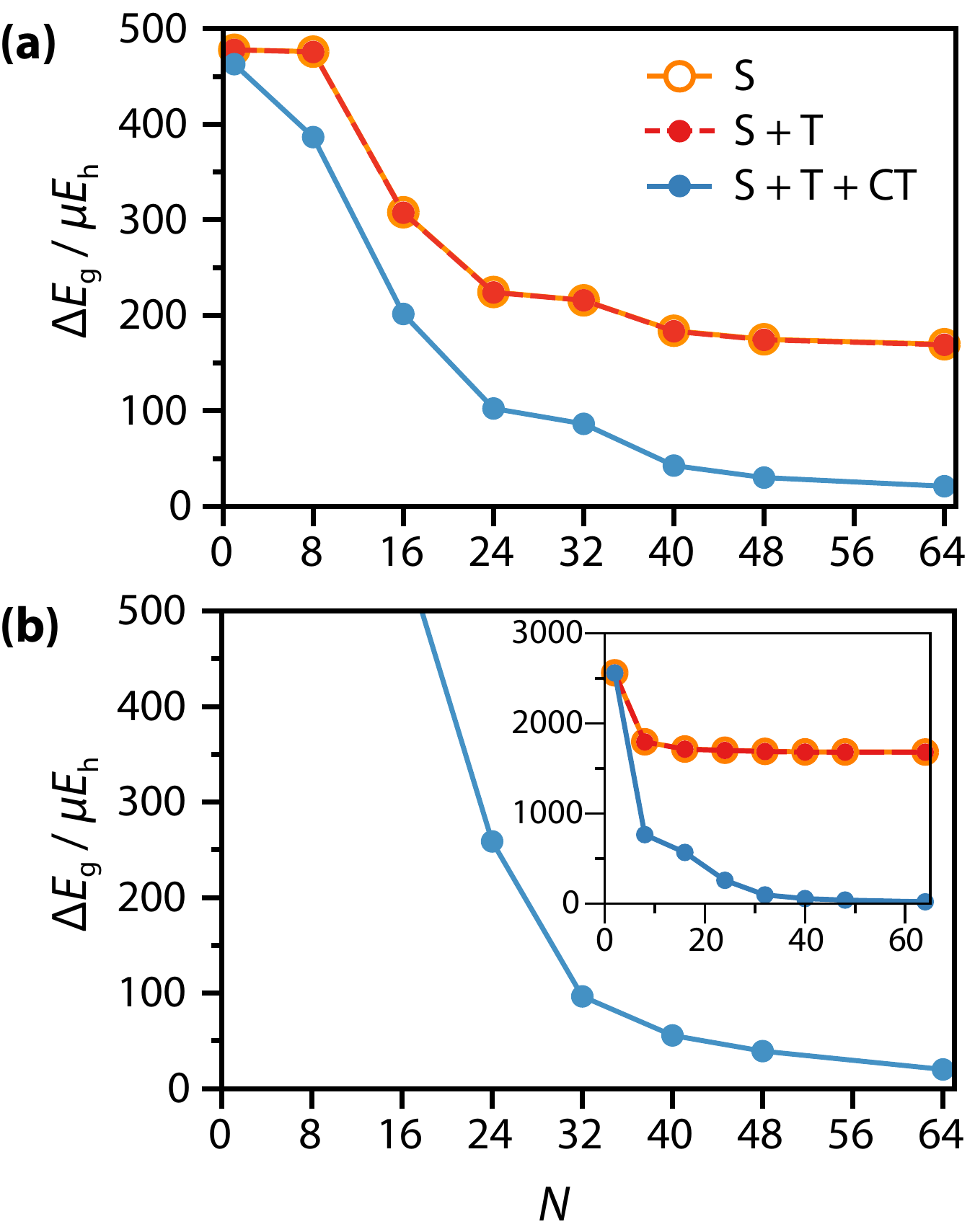}
\caption{\label{fig:c_dimers} 
Errors in the ASD-CASSCF ground-state energies as functions of the number of states in each charge and spin sector
in the ASD expansion ($N$) for (a) 4-(2-naphthylmethyl)-benzaldehyde, and (b) [$3_6$]cyclophane.
}
\end{figure}

\begin{figure}[t]
\includegraphics[width=0.45\textwidth]{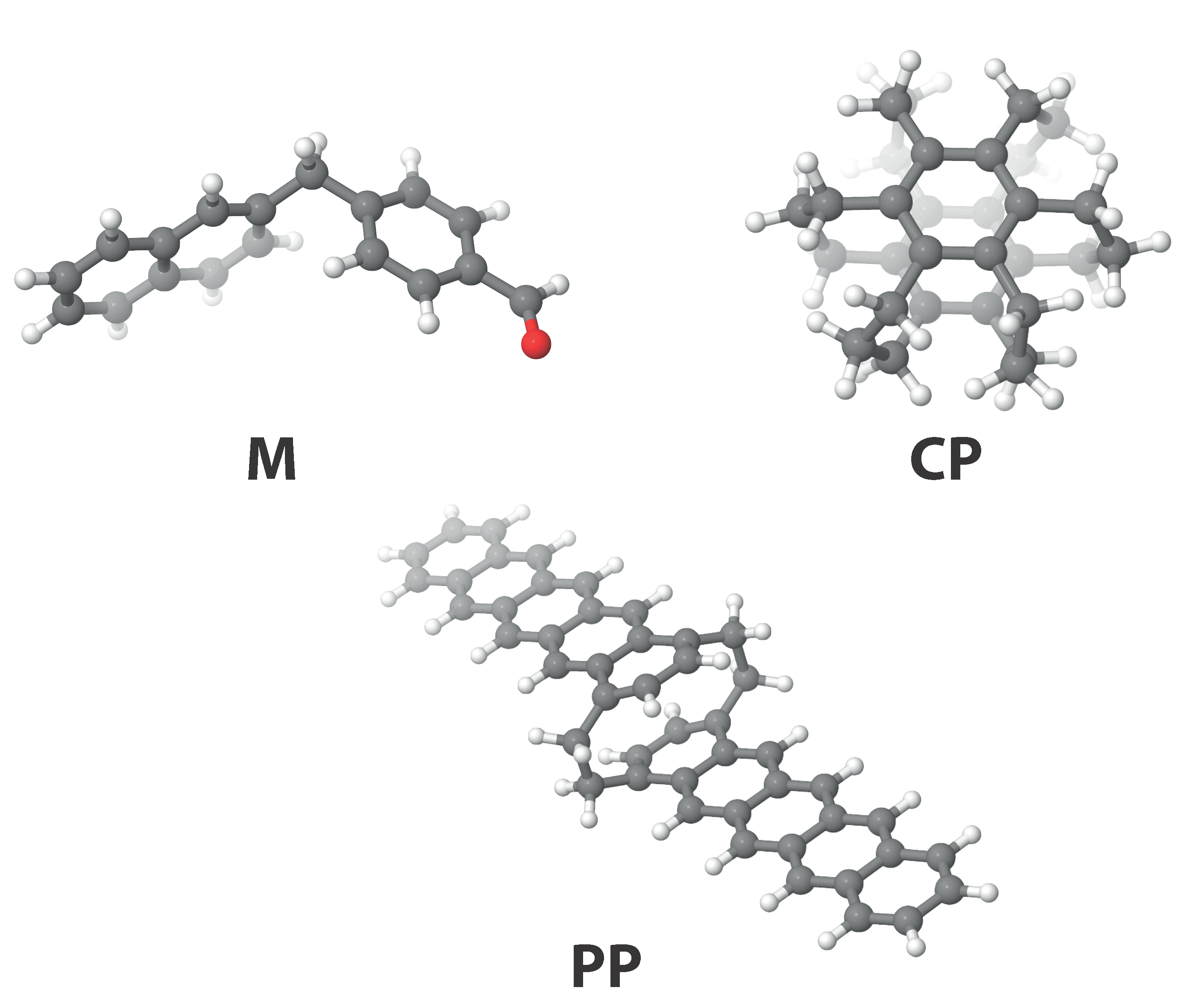}
\caption{\label{MandCP} Structures of 4-(2-naphthylmethyl)-benzaldehyde (\textbf{M}), [$3_6$]cyclophane (\textbf{CP}),
and {\it anti}-[2.2](1,4)pentacenophane (\textbf{PP}).
}
\end{figure}

Next, the energies of 4-(2-naphthylmethyl)-benzaldehyde (\textbf{M}) and [$3_6$]cyclophane (\textbf{CP})
are presented in Fig.~\ref{fig:c_dimers} as functions of the number of states in the ASD decomposition (see Refs.~\onlinecite{Closs1988JACS}, \onlinecite{Closs1989JACS}, and \onlinecite{Nogita2004JACS} for experimental studies on these molecules).
The geometries were optimized by density functional theory using the B3LYP functional\cite{Becke1993JCP} and the 6-31G* basis set.\cite{Hariharan1973TCA}
The optimized structures of these molecules are shown in Fig.~\ref{MandCP}.
The def2-SVP basis set\cite{Weigend2005PCCP} was used in the ASD calculations.
In this example, ASD-CASCI cannot be used for these molecules; therefore, only the results with ASD-CASSCF are shown.
The active space for \textbf{M} contains all the $\pi$-electrons from benzaldehydyl and naphthyl groups to form CAS(18,18),
which is decomposed to a product space of $(8 \pm q,8)$ and $(10 \mp q,10)$ where $q$ ($=0, 1, \cdots$) is the number of electrons transferred in the dimer basis.
The construction of ASD subspaces for \textbf{CP} is analogous to that for a benzene dimer above.
For \textbf{CP}, the CASSCF energy ($-1156.239888\,E_\mathrm{h}$) is used as a reference, while
the ASD-CASSCF energy with $N=128$ ($-764.327532\,E_\mathrm{h}$) is used for \textbf{M} since the exact dimer calculations with CAS(18,18) calculation are not feasible.
The large $\pi$-overlap of \textbf{CP} due to constrained geometry yields the slower convergence with respect to the number of states
and the pronounced polarization contributions to the energies in comparison to \textbf{M};
however, with $N = 32$ energies for both molecules are converged to less than 0.1 m$E_\text{h}$.
The errors presented in Fig.~\ref{fig:c_dimers} are dominated by that from the ASD approximation to the active CI, and not by variations of orbitals with respect to $N$.
To verify this, we computed the ground-state energy using ASD-CASCI with orbitals obtained by $N=8$ ASD-CASSCF.
The convergence was found to be almost identical to that of ASD-CASSCF, attesting that the change of orbitals is a minor contribution to the ASD errors.

We then turn to the triplet energy transfer processes of the donor-bridge-acceptor systems pioneered by Closs and co-workers,\cite{Closs1988JACS, Closs1989JACS}
which have been extensively studied theoretically by Subotnik and co-workers.\cite{Subotnik2010JPCA, Alguire2014JPCA, Landry2014JCTC}
The donor and acceptors are benzaldehyde and naphthalene, respectively, and the bridge is either cyclohexane or {\it trans}-decalin, each of which is rigid and saturated.
In Marcus theory, the rate of triplet energy transfer is given by\cite{Nitzanbook}
	\begin{align}	
	k = \frac{2\pi}{\hbar} |H_{IF}|^2 \sqrt{\frac{1}{4 \pi k_B T \lambda}} \exp\left( -\frac{(\lambda + \Delta G^0)^2}{4 \lambda k_B T}  \right), \label{eq:rate} 
	\end{align}
where $\Delta G^0$ is the driving force, $\lambda$ is the reorganization energy, and $H_{IF}$ is the diabatic coupling between the two diabatic states, defined in Eq.~\eqref{eq:dc}.
The triplet energy transfer is mediated by charge-transfer states, whose contributions are 
effectively treated using the quasi-degenerate second-order perturbation theory in our model. The diabatic coupling now reads\cite{Parker2014JCTC}
	\begin{align}
	\tilde{H}_{IF} = H_{IF} + \frac{1}{2} \sum_{Z \notin \mathrm{model}} \left( \frac{H_{IZ} H_{ZF}}{E_I - E_Z} + \frac{H_{IZ} H_{ZF}}{E_F - E_Z}  \right).
	\end{align}
Here, the second term refers to the perturbative correction to the direct coupling, and the sum runs over all diabatic dimer states $Z$ that are not included in the model spaces.
We used geometries optimized by B3LYP with the 6-31G* basis set.\cite{Becke1993JCP,Hariharan1973TCA}
Two triplet states were included in the state averaging, and the def2-SVP basis set\cite{Weigend2005PCCP} was used in the ASD calculations.
The active spaces are the same as that for \textbf{M} [i.e., decomposed CAS(18,18)], and 64 states were included in each charge and spin sector of the ASD expansion.
Table \ref{tab:coupling} compiles the calculated diabatic coupling and rates from the state-averaged ASD-CASSCF orbitals.
The mediated coupling values are tabulated; corresponding direct coupling values are 0.074, 0.040, 0.0078, and 0.00033 in meV for 1,3-C-ee, 1,4-C-ee, 2,7-D-ee, and 2,6-D-ee, respectively.

	\begin{table}
	\caption{Diabatic coupling $|H_{IF}|$ (in meV) and triplet energy transfer rate $k$ (in s$^{-1}$) for the Closs systems. The factors in the rate expression other than the diabatic coupling
      were taken from Ref.~\onlinecite{Subotnik2010JPCA}.
	}
	\label{tab:coupling}
	\begin{ruledtabular}
	\begin{tabular}{ccccccc}
	\multicolumn{2}{c}{} & \multicolumn{2}{c}{ASD-CASSCF} & \multicolumn{2}{c}{CIS-Boys\footnotemark[1]} \\	
	\cline{3-4} \cline{5-6}
    Molecule & $n_\sigma$\footnotemark[2] & $|H_{IF}|$ &  $k$ & $|H_{IF}|$ &  $k$ & $k_\mathrm{expt}$\footnotemark[3]\\ 
	\hline   
	1,3-C-ee 	& 3 &	0.863 &		$7.0 \times 10^9$ &	1.5\,\,\,\,\,\, &	$2.1 \times 10^{10}$ &	$7.7 \times 10^9$  \\
	1,4-C-ee 	& 4 &	0.363 &		$1.6 \times 10^9$ &	0.56\,\,\, &	$3.9 \times 10^9$ &	$1.3 \times 10^9$ \\
	2,7-D-ee 	& 5 &	0.076 &		$6.9 \times 10^7$ &	0.17\,\,\, &	$3.5 \times 10^8$ &	$9.1 \times 10^7$\\
	2,6-D-ee 	& 6 &	0.010 &		$1.1 \times 10^6$ &	0.020 &	$5.0 \times 10^6$ &	$3.1 \times 10^6$\\
   	\end{tabular}
	\end{ruledtabular}
	\footnotetext[1]{Values based on configuration interaction singles diabatized by Boys localization, taken from Ref.~\onlinecite{Subotnik2010JPCA}.}
	\footnotetext[2]{Number of $\sigma$-bonds between the donor and acceptor.}
	\footnotetext[3]{Experimental values taken from Ref.~\onlinecite{Closs1989JACS}.}
	\end{table}

\begin{figure*}[tb]
\includegraphics[width=0.9\textwidth]{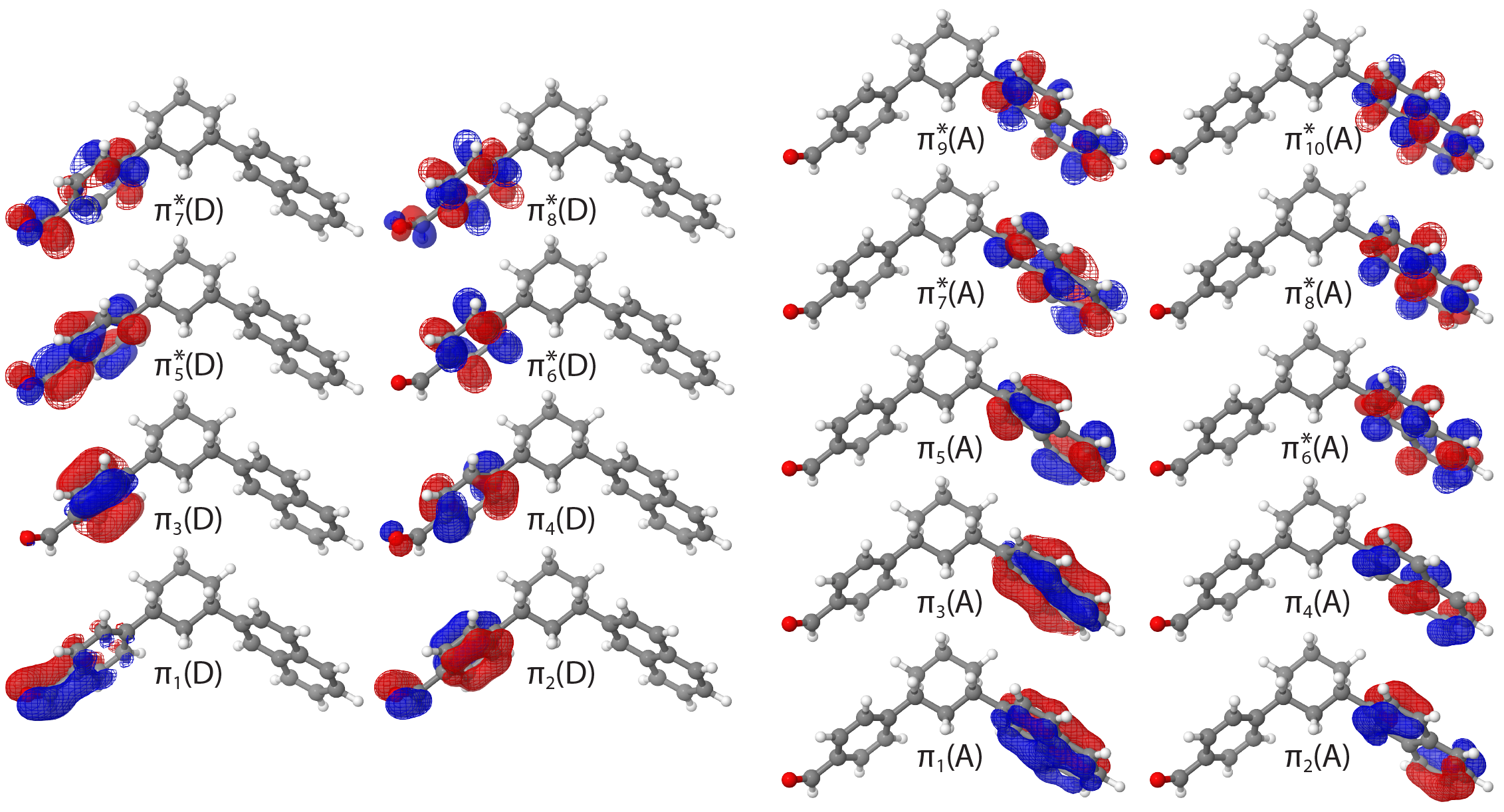}
\caption{\label{fig:mo_diff} 
Initial (mesh) and converged (filled) active semi-canonical orbitals for 1,3-C-ee from state-averaged ASD-CASSCF.
}
\end{figure*}

Combining the diabatic coupling elements with the prefactors reported by Subotnik and co-workers,\cite{Subotnik2010JPCA}
the triplet energy transfer rates were calculated using Eq.~\eqref{eq:rate}.
Our coupling values are roughly half of those obtained by Subotnik and co-workers based on the diabatization of configuration interaction singles (CIS) adiabatic states via Boys localization.
The calculated rates are in good agreement with the experiments.
For more consistent comparison to experimental results, one needs to determine the prefactors using ASD-CASSCF.
It is also noted that  we obtained the diabatic coupling of 0.54 meV for \textbf{M}, one of the Closs systems, in good agreement with 0.56 meV by Subotnik,
although the Condon approximation is not valid for this molecule owing to the flexible methylene bridge.

Figure~\ref{fig:mo_diff} visualizes the initial and the optimized active orbitals for 1,3-C-ee to
show that the diabatic nature of dimer product functions is retained during the orbital optimization procedure including active--active rotations [see Eq.~\eqref{eq:grad}]. 
The shapes of the $\pi^*$ orbitals slightly changed during the optimization, while the $\pi$-orbitals remained almost identical.
The locality of the optimized orbitals should be ascribed to the fact that the ASD expansion [Eq.~\eqref{eq:wf}] is most accurate with localized ASD subspaces.
The delocalization of the active orbitals to the bridge site is suppressed in these examples since the active space is constructed with only $\pi$-orbitals and the $sp^3$-hybridized bridging atom prevents the mixing of $\pi$- and $\sigma$-orbitals.
This allows us to construct diabatic models using the ASD-CASSCF method.

Finally the model Hamiltonians for hole and electron transfer processes of {\it anti}-[2.2](1,4)pentacenophane ({\bf PP}) were calculated using ASD-RASSCF with $N=8$.
{\bf PP} has recently been synthesized as a novel pentacene derivative.\cite{Bula2013Angew}
The charge (especially hole) mobilities of large acenes are of interest for photovoltaic applications.\cite{Anthony2008ACIE}
We used a molecular structure optimized by density functional theory (B3LYP/6-31G*).\cite{Becke1993JCP,Hariharan1973TCA}
The S, T, and CT subspaces were included in the ASD expansion,
in which RAS(7,4,7)[2,2] (7, 4, and 7 orbitals were assigned to the RAS {I}, {II}, {III} spaces for each monomer, respectively; up to 2 holes and 2 particles were allowed in the RAS {I} and {III} spaces) was used for monomer states.
We fixed the monomer CI coefficients after five iterations and 
recomputed the ASD-CI energies using the optimized orbitals (see Sec.~\ref{theorysec}).
This procedure was iterated until consistency between orbitals and monomer CI coefficients was attained.
The diabatic couplings obtained from ASD-RASSCF are 125.1 and 11.6~meV
for hole and electron transfer processes, respectively.

\section{Conclusions}

We derived and implemented an orbital optimization method
in the framework of ASD, in which a dimer wave function is expanded in products of monomer functions.
The active--active rotations between ASD subspaces are included in the optimization.
This not only removes the ambiguity in the choice of ASD subspace orbitals but also extends the applicability of ASD to covalently linked dimers.
The energies computed using both ASD-CASSCF and ASD-RASSCF converge rapidly with respect to the number of states in the ASD expansion.
Since orbital optimization preserves the diabatic nature of the dimer basis functions,
we were able to compute orbital-optimized model Hamiltonians for electron, hole, and triplet transfer processes.
Work toward incorporating dynamical correlation on the basis of the ASD-CASSCF and ASD-RASSCF methods is in progress.
The ASD-CASSCF and ASD-RASSCF methods also provide smooth potential energy surfaces and can be used to study reaction dynamics.
All the programs developed in this work are available in the open-source {\sc bagel} package.\cite{bagel}

\begin{acknowledgments}
This work has been supported by Office of Basic Energy Sciences, U.S. Department of Energy (Grant No. DE-FG02-13ER16398).
\end{acknowledgments}

\end{document}